# NUMERICAL STUDY ON TEMPERATURE VARIATIONS OF SUPERHEATED STEAM FLOWING THROUGH A REGULATION VALVE

**Zhe-hui Ma**
Institute of Advanced Equipment, Zhejiang University
Hangzhou, China

**Hang-ye Zhang**
Institute of Advanced Equipment, Zhejiang University
Hangzhou, China

**Chuang Liu**
Institute of Advanced Equipment, Zhejiang University
Hangzhou, China

**Ming Zhang**
Hangzhou Worldwides Valves Co., Ltd
Hangzhou, China

**Jin-yuan Qian**
Institute of Advanced Equipment, Zhejiang University
Hangzhou, China

## ABSTRACT

Superheated steam is widely employed in various energy systems, particularly in power plants, chemical industries, and other applications where high-temperature and high-pressure steam is essential for efficient energy conversion and process control. In these systems, regulation valves are crucial components that control the flow of steam, adjusting its pressure and temperature to ensure safe and efficient operation. Accurate understanding and prediction of temperature variations within regulation valves are essential for optimizing their performance and improving the overall system efficiency. This study investigates the temperature variations of superheated steam flowing through a regulation valve using computational fluid dynamics (CFD) simulations combined with Proper Orthogonal Decomposition (POD) techniques. The primary goal is to analyze the behavior of steam within the valve and to identify the underlying flow structures that contribute to temperature fluctuations, particularly in the throttling regions where steam undergoes rapid changes in pressure and temperature. The numerical model is first established and validated through experiments conducted at operating conditions of 2 MPa pressure and 574°C steam temperature. These conditions are representative of typical industrial steam systems, allowing for a realistic simulation of the phenomena. The analysis begins with an examination of the internal flow field parameters, including temperature and pressure, to understand the overall fluid dynamics within the valve. Special focus is given to the throttling regions, where significant variations in both temperature and pressure occur due to the Joule-Thomson effect. This phenomenon results in temperature changes as the steam passes through the restriction of the valve, which can have important implications for system performance and energy efficiency. To gain further insights into the complex three-dimensional flow patterns, POD is applied to reduce the dimensionality of the CFD results. This approach involves extracting the most dominant flow modes from the high-dimensional data generated by CFD simulations. The original 3D CFD data is imported into MATLAB, where resampling is performed to prepare the data for POD analysis. Singular Value Decomposition (SVD) is then employed to extract the dominant modes that capture the key flow structures responsible for heat transfer and temperature fluctuations. The POD analysis reveals that the most influential modes are associated with regions of high turbulence intensity and significant temperature gradients, which are critical to the thermal performance of the steam flow through the regulation valve. The application of POD to 3D CFD results represents a novel approach, particularly for complex fluid flow models such as steam flow through regulation valves. This method offers a tool for simplifying the analysis of intricate flow phenomena, enabling the identification of key flow features that influence heat transfer. The insights gained from this study have practical implications for the design and optimization of temperature and pressure regulation valves in energy systems, providing a theoretical foundation for enhancing the efficiency and reliability of these systems.

## NOMENCLATURE

| | |
|---|---|
| $u$ | Physical quantity |
| $\phi$ | Spatial mode of POD |
| $a$ | Coefficient of POD |
| $k$ | Order of modes |
| $A_z$ | Throttling area |
| $p_i$ | Pressure of inlet |
| $p_o$ | Pressure of outlet |
| $\gamma$ | Adiabatic exponent |

## 1. INTRODUCTION

Steam is widely used in both carbon-intensive [1] and green [2–4] energy engineering for power generation. In power plants, as the superheated steam produced from generators cannot be taken into application directly, regulation valves are needed to reduce the temperature and pressure of the steam [5]. Thus, these valves are crucial components that control the steam flow, ensuring safe and efficient operation [6–8]. Therefore, understanding and prediction of temperature variations within them are essential for improving the overall system efficiency.

Temperature control is one of the key functions of regulation valves [9]. Therefore, numerous studies have been done to investigate the temperature variation inside valves. These studies can help optimize the design of valves and improve the efficiency of industrial systems. Si et al. [10] analyzed the impact of water spray on temperature variations within a temperature and pressure reducing valve, finding that the evaporation of smaller droplets is more effective in controlling steam temperature and reducing thermodynamic losses. Hong et al. [11] studied how oil temperature affects the flow characteristics of a hydraulic valve, finding that higher oil temperature lead to reduce viscosity and axial force on the valve core, which could improve control precision. Essam et al. [12] took temperature effects into valve flow modeling, leading to more accurate predictions. It was found that the machine learning model significantly reduces prediction errors, while the simplified analytical model struggles with data outside its tuned range. Ngwa et al. [13] explored heat transfer in a pneumatic control valve, noting that airflow turned icy in high-speed areas. It concluded that CFD simulations, validated by experiments, accurately captured these temperature drops, aiding in valve design optimization. Monika et al. [14] conducted numerical simulations to study the cooling of electric vehicle batteries. They found that a multi-stage Tesla valve cooling system improves heat transfer by using reverse flow, resulting in increased pressure drop. Based on this discovery, the design of the multi-stage Tesla valve was optimized for better cooling performance. Singh et al. [15] analyzed how different throttle valve shapes affect temperature and velocity in a vortex tube. It found that the truncated valve shape resulted in an 11.52% increase in cold outlet temperature compared to the square shape and an 8.98% increase over the conical shape, with the best coefficient of performance and isentropic efficiency.

In addition to simulation, Proper Orthogonal Decomposition (POD) has also been successfully applied to analyze flow fields in engineering for a long time[16–18]. Hu et al. [19] employed POD to analyze the main flow fields of Tesla valves. It was found that the diode performance is optimal when the angle θ is between 70° and 80°, with the diode property being primarily controlled by separation bubbles near the left junction. Hoffmann et al. [20] used POD to analyze flow fields and found that valve lift significantly affects flow structures, while mass flow variation does not. POD revealed that turbulent kinetic energy increases with mass flow and decreases with valve lift, leading to high-energetic structures near the jet and low-energetic structures throughout the flow field. Then the study done later [21] showed that large, high-energy flow structures dominate near the jet, while smaller, low-energy structures are evenly spread across the flow field. Vijay et al. [22] used POD to analyze flow fields in pulsejet engines, finding that engine geometry influences the strength of two main vortical systems. They also discovered that a diverging nozzle flare leads to distinct starting vortices, related to the fluidic piston effect and Helmholtz resonance. POD analysis by Liu et al. [23] revealed that modulating fuel with a high-speed valve can significantly dampen flame oscillations at a 40° phase, yet intensifies them at a 220° phase, highlighting the impact of phase differences on flame behavior.

Although lots of studies have been done to understand the complex dynamics within regulation valves, recent research shows limitations on the application of POD to complex geometric models like valves, and there is still uncertainty regarding the detailed mechanisms of superheated steam flow within the valves.

To address these limitations, this paper establishes a numerical model of a regulation valve used in a power plant and validates its accuracy with an experiment platform. The behavior of steam within the valve is analyzed. POD method is introduced to identify the underlying flow structures that contribute to temperature and pressure variations.

## 2. MATERIALS AND METHODS
### 2.1 Physical model and numerical method

The section view of the temperature and pressure regulation valve is shown in Figure 1, which mainly consists of two sleeves, a valve seat, and a valve body. Since only the fluid domain of the valve is studied, components unrelated to the fluid domain are removed, including the valve stem and valve deck. The nominal diameter of the valve is DN400. The diameter of holes on the inner sleeve is 7 mm, while that on the outer sleeve is 6 mm. When the regulation valve is in operation, steam with a pressure of 2MPa and a temperature of 573 °C enters from the inlet, flows through two sleeves, and then flows out from the outlet. The parameters of steam at the outlet vary with the opening height of the valve, which is the lifting height of the inner sleeve.

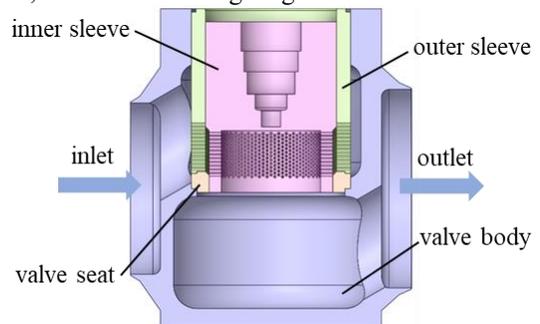

**FIGURE 1:** GEOMETRY MODEL OF THE REGULATION VALVE

The Mesh of the fluid domain is shown in Figure 2. Considering computational efficiency and symmetry of the model, half of the fluid domain is used for simulation. To ensure that the steam flow is fully developed, the length of the inlet pipe is set as 2.5m, and the length of the outlet pipe is set as 5m. The size of the grids at the holes of sleeves is a minimum of 1.25 mm. This is due to the complexity of nearby flows and the small dimensions of the holes.

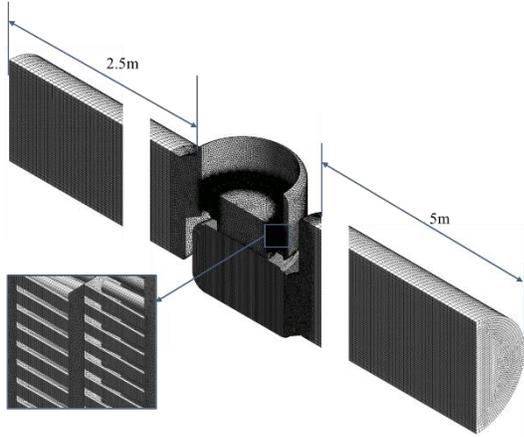

**FIGURE 2:** MESH OF FLUID DOMAIN IN THE REGULATION VALVE

The mesh-independent verification is conducted as shown in Table 1. In CFD simulation, the inlet pressure, inlet temperature, and outlet pressure are 2MPa, 574 °C, and 1.3 MPa respectively. As the mesh number increases, the relative deviation of the outlet flow rate of the regulation valve becomes lower and lower. When the mesh number reaches 9.43 million, the relative deviation is less than 1%. Considering both computational error and efficiency, the mesh method of about 9.43 million elements is selected.

**TABLE 1:** MESH INDEPENDENCE ANALYSIS

| Mesh number ($10^4$) | Outlet flow rate (kg/s) | Relative deviation |
|---|---|---|
| 441 | 14.29 | - |
| 723 | 15.06 | 5.13% |
| 943 | 15.15 | 0.61% |
| 1302 | 15.21 | 0.37% |

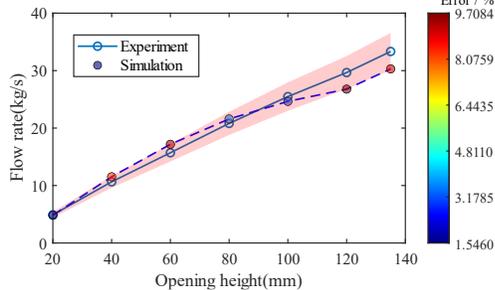

**FIGURE 3:** NUMERICAL METHOD VALIDATION

In order to validate the accuracy of simulation results, the simulation data is compared with the experiment data. The experiment and simulation are both conducted with an inlet pressure of 2MPa, an outlet pressure of 1.3MPa, and a steam temperature of 574 °C. As shown in Figure 3, the maximum deviation between simulation and experiment is 9.7084%, which is less than 10%. This means the CFD simulation is relatively accurate and has reference value for practical engineering.

The experiment platform components are shown in Figure 4, including the pipeline system and regulation valve. As the temperature of steam is extremely high (574 °C), the pipelines, valves and flowmeter are wrapped in insulation material. It should be noted that, since the experiment data used in Figure 3 is the outlet flow rate of the regulation valve, the primary source of measurement uncertainty in the data is the flowmeter. The maximum error of the V-cone flowmeter used in the experimental platform is 0.5%, and its uncertainty is ± 0.118% ~ ± 0.203%.

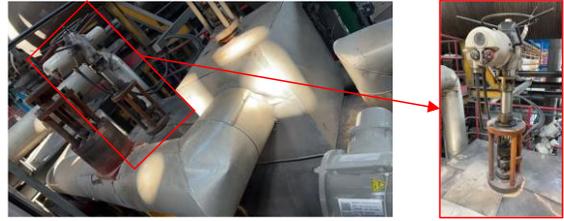

**FIGURE 4:** EXPERIMENT PLATFORM AND VALVE

### 2.2 POD method

POD is an attempt to decompose turbulent fluid motion into a set of flow modes [24]. Each mode captures a portion of the total energy. It is often computed using the singular value decomposition (SVD) [25]. For time-series data [26], POD can decompose a physical quantity in the flow field into the product of time coefficients and spatial modes so that:

$$u(X,t) = \sum_{k=1}^{\infty} a_k(t)\phi_k(X) \quad (1)$$

where $u(X,t)$ is the physical quantity, $a_k(t)$ is the time coefficient, $\phi_k(x)$ is the spatial mode, $t$ is the time, $X$ is the position vector, $k$ is the order of the mode. While in this study, the spatial structure of the flow inside the valve is the focus, the temporal evolution of the flow is not considered. Therefore, the POD in this study is based on simulation data of the flow field at different opening heights. That means the geometric models of the various simulation data used for POD are different. This method has been proven workable in some studies [27,28].

The analysis of this study mainly focuses on spatial modes $\phi_k(x)$. Due to the property of POD, these spatial modes are orthogonal [29], which means:

$$\iiint_X \phi_{k1}(X)\phi_{k2}(X)dx = \begin{cases} 1, k_1 = k_2 \\ 0, k_2 \neq k_2 \end{cases} \quad (2)$$

Another property of POD is that the energy of each spatial mode is different [30]. The energy of spatial modes of velocity, pressure, and temperature is shown in Figure 5. When the total energy of the first few modes is above 90% [31], it can be considered that these modes are sufficient to represent the entire flow. As shown in Figure 5 (a), the first mode has the most of energy, specifically around 64.84% of velocity, 75.68% of pressure, 74.20% of density, and 44.46% of temperature, respectively.

3© 2025 by ASME

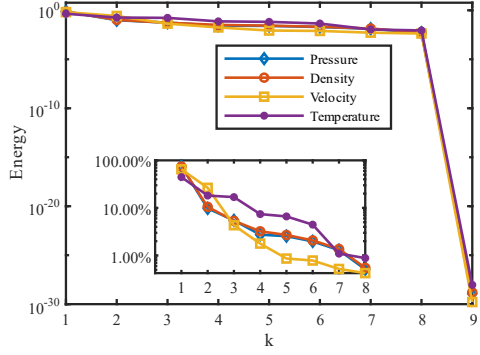

(a) Energy of spatial modes

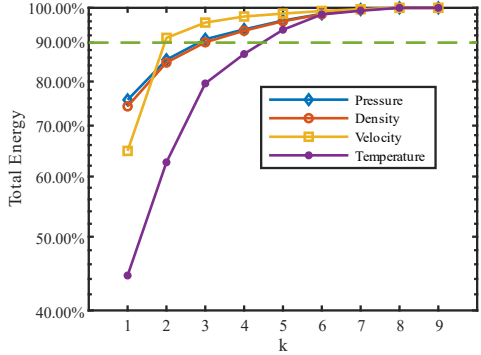

(b) Total energy of spatial modes
**FIGURE 5:** ENERGY OF SPATIAL MODES

Before conducting POD analysis, it is necessary to validate the accuracy of the POD method. The reconstruction result based on the first 4 modes is compared with the experiment data in Figure 6. The maximum deviation between POD and experiment is 13.0277%. That means the POD method is relatively accurate and has reference value for practical engineering. It is worth noting that the error of POD is larger than that of simulation. The source of error should be resampling.

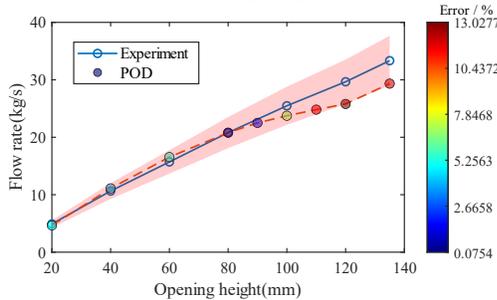

**FIGURE 6:** POD METHOD VALIDATION

## 3. RESULTS AND DISCUSSION

The results are presented in two sections. Firstly, the internal flow field parameters are analyzed. Special focus is given to the throttling regions, where significant variations in both temperature and pressure occur. Secondly, POD is applied to reduce the dimensionality of the CFD results and extract the dominant spatial modes. These modes are analyzed to capture the key flow structures responsible for heat transfer and temperature variations.

### 3.1 Internal flow field analysis

The temperature variation of superheated steam in the regulation valve is shown in Figure 7. Superheated steam flows into the inlet pipeline, passes through the throttle holes on the sleeves, and then flows into the outlet pipeline from the valve chamber. The main temperature variation occurs in the throttle holes and the valve chamber. Among them, the temperature variation in the throttle hole is due to a sudden decrease in the flow area. The temperature variation in the valve chamber is due to the convergence of the fluid. Due to the Joule Thomson effect, the temperature decreases at the throttle hole of the sleeve, with a minimum temperature of about 530 °C. Figure 7 shows that the minimum temperature at the symmetry of sleeves under different valve opening heights does not change significantly. That is because in simulation and experiment, the inlet and outlet pressures are set as 2 MPa and 1.3 MPa, respectively.

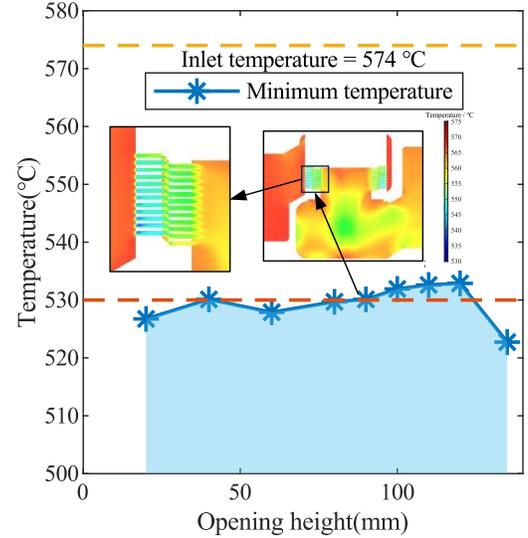

**FIGURE 7:** TEMPERATURE VARIATION

The flow of high-pressure superheated steam through throttling components can be regarded as an isoenthalpy throttling expansion process. Thus, only the temperature variation caused by Joule Thomson effect is considered[32]. That means:

$$\Delta T = \int_{p_1}^{p_2} \mu_{Jt} dp \qquad (3)$$

where $\Delta T$ is the temperature variation on the throttling components, $p_1$ is the inlet pressure of the throttling components, $p_2$ is the outlet pressure of the throttling components, $\mu_{Jt}$ is the Joule Thomson coefficient. In this study, the inlet and outlet pressures are fixed. Therefore, the minimum temperatures at the throttle holes are roughly the same at different opening heights.

About the temperature variation, it is worth noting that from the temperature distribution on the symmetry plane in Figure 7, it can be observed that the minimum temperature on the outer sleeve is lower than that on the inner sleeve. That is to say, the pressure drop of the first stage decompression should be higher than that of the second. Only if such pressure drop condition is met, the temperature distribution in Figure 7 is reasonable. Thus, it is necessary to analyze the pressure variation inside the regulation valve.

The pressure variation of superheated steam flowing through the regulation valve at a valve opening height of 120 mm is shown in Figure 8. Steam with a pressure of 2 MPa enters from the inlet. After flowing through the outer sleeve, the pressure of steam decreases from 2 MPa to approximately 1.53 MPa. After flowing through the inner sleeve, the pressure of steam decreases from 1.53 MPa to approximately 1.42 MPa. Subsequently, due to the convergence of steam, there is a slight increase in pressure inside the valve chamber. Finally, due to the corner between the valve chamber and the outlet pipe, the pressure of steam decreases to 1.3 MPa.

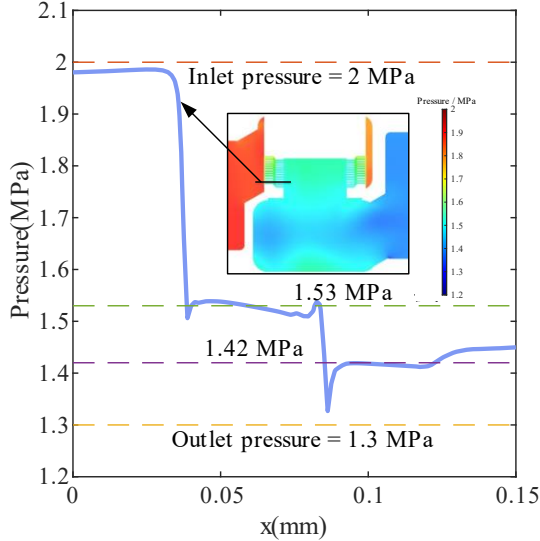

**FIGURE 8:** PRESSURE VARIATION

In the design of temperature and pressure regulation valve, the throttling area can be calculated by:

$$A_z = \frac{W}{C_s \sqrt{2g \frac{\gamma}{\gamma - 1} p_i \rho_i \left[\left(\frac{p_o}{p_i}\right)^{\frac{2}{k}} - \left(\frac{p_o}{p_i}\right)^{\frac{k+1}{k}}\right]}} \quad (4)$$

where $A_z$ is the throttling area, $W$ is the mass flow rate, $C_s$ is a constant coefficient determined by the kind of gas, $g$ is the gravitational acceleration, $\gamma$ is the adiabatic exponent, $p_i$ is the pressure of inlet, $\rho_i$ is the density of steam at inlet, $p_o$ is the pressure of outlet. This equation means that when the throttling area remains constant, the higher the inlet pressure, the higher the magnitude of decompression. For this regulation valve, the difference in throttling area between the outer sleeve and the inner sleeve is not significant. And in any case, the inlet pressure of the first stage decompression is higher than that of the second. That's why the pressure drop at the outer sleeve is higher than that at the inner sleeve in Figure 8.

### 3.2 POD modes analysis

Based on section 2.2, the number of modes analyzed is selected as 4. To reveal the spatial features of underlying flow structures, the POD modes obtained from the symmetry of pressure fields at a valve opening height of 120 mm are presented in Figure 9. For the whole fluid domain, the energy of these four modes accounts for 75.68%, 9.77%, 5.47%, and 2.74% of the total energy, respectively. In all 4 modes, the pressure variations at the throttle holes are quite significant. In mode 1, the negative pressure on the outer sleeve is very obvious, lower than the pressure on the inner sleeve. This is consistent with the analysis in section 3.1. With the change of valve opening height, mode 1 will influence the pressure distribution on the sleeves significantly, especially at the middle part of the sleeves.

Another point worth noting is that in mode 2, 3 and 4, there is a significant difference in pressure distribution between the upper and lower parts of the sleeves. This might be related to the angle of the inlet channel. After entering the valve from the inlet pipe, superheated steam needs to flow through an inclined channel to enter the valve chamber. This means that the pressure loss of superheated steam flowing through the lower part of the sleeves is smaller. Therefore, the pressure on the lower part of the sleeve is higher than that on the upper part.

Apart from the sleeves, there is also a slight fluctuation in pressure in the valve chamber. Obviously, fluid convergence does have an impact on pressure, but it is particularly small compared to the effect of throttle holes on the sleeves.

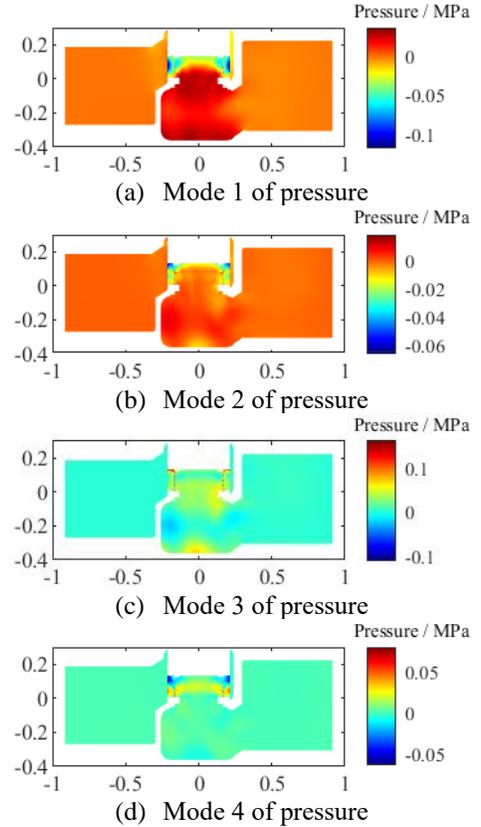

**FIGURE 9:** MODES OF PRESSURE

In order to further reveal the temperature variations in the regulating valve, the POD modes obtained from the symmetry of temperature fields at a valve opening height of 120 mm are presented in Figure 10. For the whole fluid domain, the energy of these four modes accounts for 44.46%, 18.18%, 16.90%, and 7.41% of the total energy, respectively.

In modes 1 and 2, the temperature at the middle part of the sleeve is higher than that of the upper and lower parts. This is different from the pressure modes. To explain this phenomenon, it is necessary to combine the Joule Thomson effect with the pressure mode shown in Figure 10 (a). After removing the mean (mode 0), the pressure difference between the inside and outside at the middle part of the sleeves is higher than that at the upper and lower part. Since mode 1 contains most of the energy, this means that the pressure variation at the middle part of the sleeves is the most severe. Therefore, under the influence of Joule Thomson effect, the phenomenon shown in Figure 10 (a) occurred.

Besides, according to these 4 modes in Figure 10, fluid convergence in the valve chamber seems to have a more significant effect on temperature than on pressure.

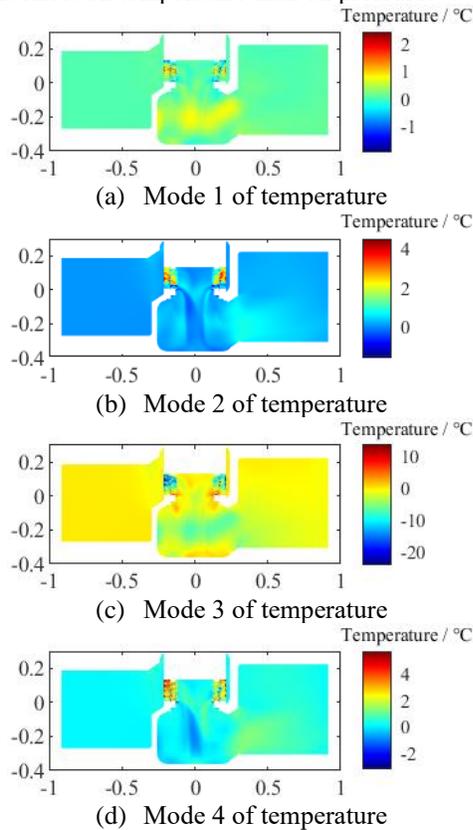

(a) Mode 1 of temperature

(b) Mode 2 of temperature

(c) Mode 3 of temperature

(d) Mode 4 of temperature

**FIGURE 10:** MODES OF TEMPERATURE

## 4. CONCLUSION AND OUTLOOK

This paper presents a numerical study on the temperature variations of superheated steam flowing through a regulation valve, employing CFD simulations combined with POD techniques. The investigation aims to analyze the internal flow field parameters, particularly focusing on the throttling regions where significant temperature and pressure changes occur due to the Joule-Thomson effect. The study also seeks to identify the underlying flow structures that contribute to these temperature fluctuations, which are essential for optimizing valve performance and enhancing system efficiency.

Based on the previous analysis, the following conclusions are drawn:

(1) It is found that the temperature of superheated steam decreases significantly when it flows through the throttle holes of sleeves. The minimum temperature is about 530 °C at throttle holes of the outer sleeve, indicating a substantial cooling effect due to the Joule Thomson effect. Notably, the temperature variation analysis reveals that the minimum temperature at the outer sleeve (first stage of decompression) is lower than that at the inner sleeve (second stage of decompression). This difference in temperature is attributed to the higher pressure drop at the first stage, which results in a more pronounced cooling effect.

(2) The application of POD identifies key flow structures within the regulation valve that significantly influence the variation of temperature and pressure. The analysis indicates that the angle of the inlet channel affects the pressure distribution across the sleeves, with the lower part of the sleeves experiencing less pressure loss due to the inclination of flow. Furthermore, the fluid convergence in the valve chamber is found to have an effect on temperature. This means that the chamber's geometry plays a role in heat transfer processes.

This study provides an insight of the internal flow dynamics in regulation valves, which is essential for optimizing their design and performance in industrial applications. Future work will focus on developing reduced-order models for accelerated computation in valve analysis, enhancing the efficiency of simulations and allowing for quicker optimization of valve performance.

## ACKNOWLEDGEMENTS

This work is supported by the National Natural Science Foundation of China (NSFC) through Grant No. 52422506.